\documentclass[a4paper,11pt]{article}

\usepackage{pos}
\usepackage{lipsum} 
\usepackage{wrapfig}
\usepackage{siunitx}
\usepackage{placeins}
\usepackage{setspace}
\usepackage{graphicx}

\newlength{\bibitemsep}
\newlength{\bibparskip}
\let\oldthebibliography\thebibliography
\renewcommand\thebibliography[1]{%
  \oldthebibliography{#1}%
  \setlength{\parskip}{\bibitemsep}%
  \setlength{\itemsep}{\bibparskip}%
}

\newcommand{\refeq}[1]{Eq.~(\ref{#1})}

\newcommand{\reffig}[1]{Fig.~\ref{#1}}

\newcommand{\refsec}[1]{Section~\ref{#1}}

\newcommand{\refref}[1]{Ref.~\cite{#1}}

\newcommand{\sibyllpre}{Sibyll~2.1}

\newcommand{\epos}{EPOS\nobreakdash-LHC}
\newcommand{\qgsjet}{QGSJet\nobreakdash-II.04}

\newcommand{\Nmu}{$N_\mu$}
\newcommand{\Nmuav}{$\langle N_\mu \rangle$}
\newcommand{\Enul}{$E$}

\newcommand{\Sone}{$S_{125}$}

\title{Multiplicity of TeV muons in extensive air showers detected with IceTop and IceCube}

\ShortTitle{Multiplicity of TeV muons in air showers with IceTop and IceCube}

\author{The IceCube Collaboration \\{\normalsize \normalfont(a complete list of authors can be found at the end of the proceedings)}\\}

\emailAdd{verpoest@udel.edu}

\abstract{

We report on an analysis of the high-energy muon component in near-vertical extensive air showers detected by the surface array IceTop in coincidence with the in-ice array of the IceCube Neutrino Observatory. In the coincidence measurement, the predominantly electromagnetic signal measured by IceTop is used to estimate the cosmic-ray primary energy, and the energy loss of the muon bundle in the deep in-ice array is used to estimate the number of muons in the shower with energies above \SI{500}{\giga\eV} (“TeV muons”). The average multiplicity of these TeV muons is determined for cosmic-ray energies between \SI{2.5}{\peta\eV} and \SI{100}{\peta\eV} assuming three different hadronic interaction models: \sibyllpre{}, \qgsjet{}, and \epos{}. For all models considered, the results are found to be in good agreement with the expectations from simulations. A tension exists, however, between the high-energy muon multiplicity and other observables; most importantly the density of GeV muons measured by IceTop using \qgsjet{} and \epos{}.

\vspace{4mm}
{\bfseries Corresponding authors:}
Stef Verpoest$^{1*}$\\
{$^{1}$ \itshape Bartol Research Institute, Dept. of Physics and Astronomy University of Delaware, Newark, DE 19716, USA}\\[4mm]
$^*$ Presenter

\ConferenceLogo{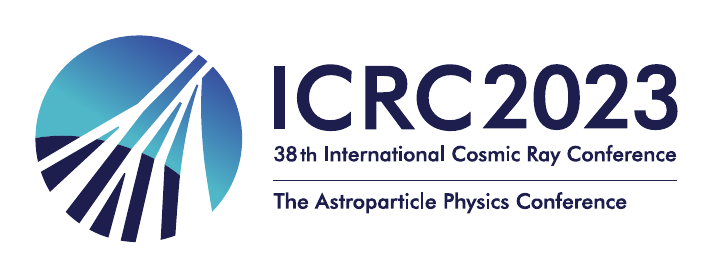}

\FullConference{The 38th International Cosmic Ray Conference (ICRC2023)\\ 26 July -- 3 August, 2023\\ Nagoya, Japan}
}

\begin{document}

\maketitle

\section{Introduction}\label{sec1}

Cosmic rays with energies above \SI{100}{\tera\eV} are studied indirectly through the extensive air showers (EAS) they initiate in the Earth's atmosphere. The interpretation of the measurements in terms of the mass of the primary nucleus relies on detailed simulations of the EAS development. A discrepancy between the predicted and observed number of muons in EAS has been reported by several experiments~\cite{EAS-MSU:2019kmv}, and is generally attributed to an incomplete description of high-energy hadronic interactions~\cite{Albrecht:2021cxw}. The uncertainties in the description of EAS development prevent an accurate determination of the cosmic-ray mass composition, an important part of the puzzle of understanding the sources of cosmic rays.

\begin{wrapfigure}{R}{0.5\textwidth}
    \centering
    \includegraphics[width=0.5\textwidth]{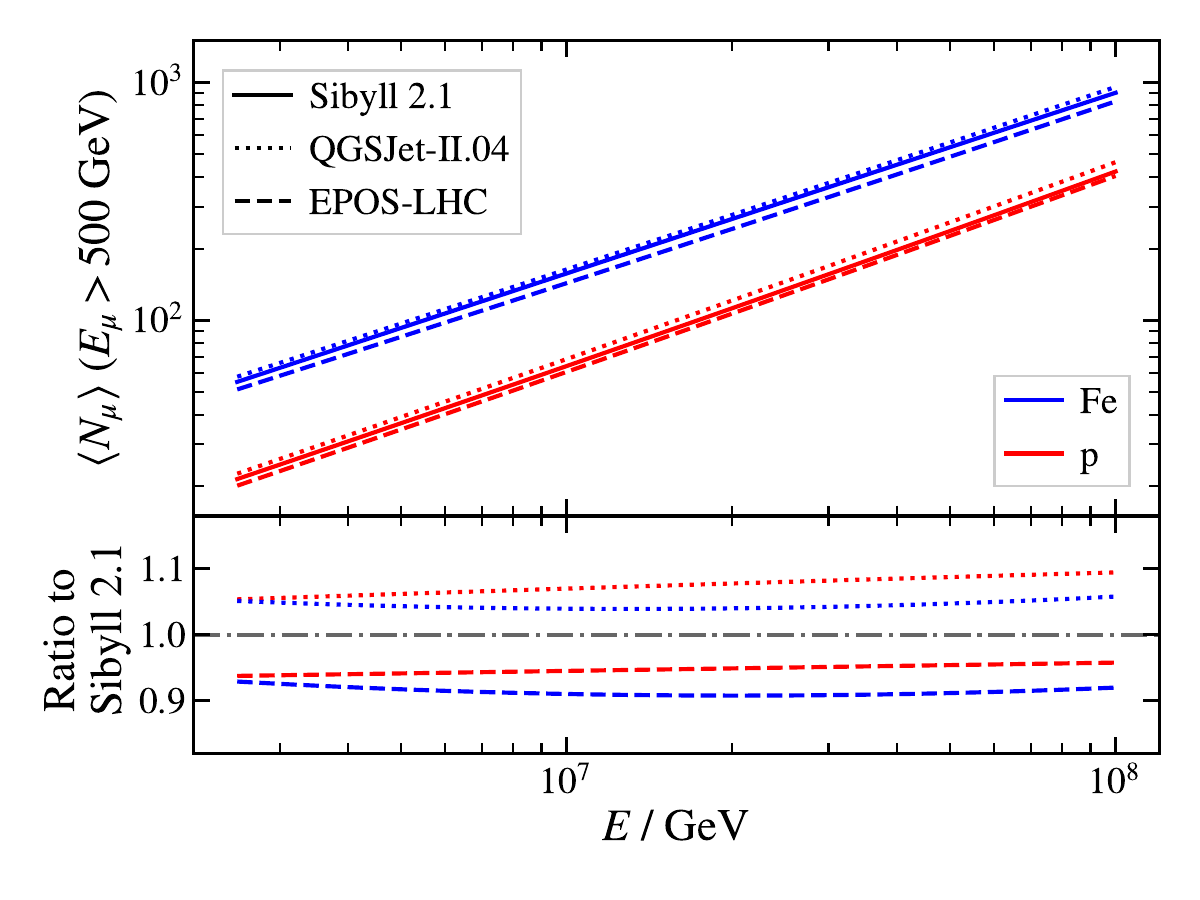}
    \caption{Average multiplicity of muons with an energy above \SI{500}{\giga\eV} in near-vertical EAS obtained from simulations for proton and iron nuclei and different hadronic models.}
    \label{fig:predictions}
\end{wrapfigure}

The IceCube Neutrino Observatory~\cite{IceCube:2016zyt} can perform unique tests of muon production in EAS with its combination of a surface air-shower array and a deep in-ice detector, detecting muons from two different kinematic regimes. The surface detector, IceTop, measures muons with energies mostly around \SI{1}{\giga\eV}~\cite{IceCubeCollaboration:2022tla}. The in-ice detector, hereafter referred to as IceCube, can only be reached by muons with energies above several hundred \si{\giga\eV}. Previous studies have demonstrated inconsistencies between air-shower observables in different hadronic interaction models~\cite{IceCube:2021ixw, Verpoest:2022ntf}.

In this work, a measurement of the high-energy muon component, referred to as TeV muons, in near-vertical air showers observed in coincidence between IceTop and IceCube is presented. More specifically, we reconstruct the average number of muons, \Nmuav{}, with an energy above \SI{500}{\giga\eV} in EAS as a function of the primary cosmic-ray energy. Predictions obtained from CORSIKA~\cite{Heck:1998vt} simulations for \Nmuav{} in the considered primary energy range of \SI{2.5}{\peta\eV} to \SI{100}{\peta\eV} are shown in \reffig{fig:predictions}. The analysis is performed using the hadronic interaction models \sibyllpre{}~\cite{Ahn:2009wx}, \qgsjet{}~\cite{Ostapchenko:2010vb}, and \epos{}~\cite{Pierog:2013ria}.

\section{IceTop \& IceCube}

The IceCube Neutrino Observatory is a multi-purpose detector located at the geographical South Pole. IceTop is the surface component of the observatory and it is situated at an altitude of about \SI{2.8}{\km} a.s.l., corresponding to an average atmospheric depth of $\sim$\SI{690}{\g\per\cm\squared}~\cite{IceCube:2012nn}. It consists of 81 stations, each comprising two ice-Cherenkov tanks, deployed on a triangular grid with a spacing of about \SI{125}{\m}. Each tank contains two Digital Optical Modules (DOMs) which detect Cherenkov photons produced by particles traversing the ice inside the tanks. IceCube consists of about 5000 DOMs inside the Antarctic ice between depths of \SI{1450}{\m} and \SI{2450}{\m}~\cite{IceCube:2016zyt}. The DOMs are deployed on vertical strings following approximately the same grid as the stations at the surface, instrumenting a total volume of about one cubic kilometer. 

IceTop detects EAS with primary energies typically between \SI{1}{\peta\eV} and \SI{1}{\exa\eV}. IceTop detects EAS with primary energies typically between 1 PeV and 1 EeV. Due to the high altitude of the array, vertical EAS are detected close to the shower maximum. IceTop signals are dominated by the electromagnetic shower component, except
for several hundred meters away from the shower axis, where muons are more prominent. Snow accumulation on top of the detectors absorbs a fraction of the particles and moves the trigger threshold up in energy over time. Muons with an energy over several hundred GeV can penetrate deeply into the ice and can be observed in coincidence in IceCube if the shower axis intersects it.

\section{Muon multiplicity analysis}

\subsection{Event selection and reconstruction}

A standard air-shower reconstruction is applied to events passing the IceTop trigger, reconstructing the direction of the shower axis, the shower core position, and the shower size~\cite{IceCube:2012nn} using only information from IceTop. This is done by fitting the measured signal sizes and times with a lateral distribution function (LDF) and shower front model respectively. The shower size \Sone{} is the signal obtained from the LDF fit at a distance of \SI{125}{\m} from the shower axis. The attenuation of the expected signal as a result of snow accumulation on the surface is accounted for in the reconstruction. Events are furthermore required to have an in-ice trigger which coincides with the expected arrival time of the high-energy muon bundle in the deep detector. The recorded in-ice signals are used to reconstruct the combined energy loss of the muons in the bundle. The reconstruction algorithm fits the deposited energy in segments of \SI{20}{\m} along a seed track~\cite{IceCube:2013dkx}, for which the reconstructed shower axis from the IceTop reconstruction is used.

Several quality cuts developed for previous analyses are applied (see \refref{IceCube:2019hmk} for details). The IceTop selection ensures that the events have a well-reconstructed direction and shower size, with their core position contained within the boundaries of the IceTop array. The IceCube selection ensures that a muon bundle signal with a successful energy-loss reconstruction is detected in IceCube. The analysis is further restricted to near-vertical showers with the reconstructed zenith angle $\cos \theta < 0.95$ or $\theta \lesssim 18^\circ$. 

This analysis uses data recorded between May 15, 2012 and May 5, 2013. For the snow coverage of this period, the selection reaches full efficiency for primary cosmic rays with energies above \SI{2.5}{\peta\eV}, which is therefore chosen as the low-energy threshold of the analysis. At high energies, the analysis is limited to \SI{100}{\peta\eV}. A total of 1216154 events passing the selection criteria are included in the analysis.

\subsection{Neural network}

A neural network is used to reconstruct two quantities for every event: the primary cosmic-ray energy, \Enul{}, and the multiplicity of muons with energy above \SI{500}{\giga\eV} in the shower, \Nmu{}, counted at the South Pole surface level. The neural network takes inputs from IceTop and IceCube. The segmented energy-loss reconstruction of the muon bundle is used as an input to a recurrent neural network layer. The output of this layer is combined with the shower size \Sone{} and zenith angle $\theta$ from the air-shower reconstruction as an input to a number of dense layers which finally return \Enul{} and \Nmu{}. \Sone{} is known to be a good estimator for \Enul{}, while the energy loss of the bundle is strongly correlated with the number of high-energy muons in the shower~\cite{IceCube:2019hmk}.

The neural network was trained on Monte Carlo simulations produced with CORSIKA v7.7300, using \sibyllpre{} as the high-energy hadronic model. The atmospheric model describes the average South Pole atmosphere in April, which is close to the yearly average. The training data includes four primary nuclei (p, He, O, Fe) and has undergone full detector simulation and processing identical to the experimental data.

\begin{figure}
    \centering
    \includegraphics[trim={0.0cm 0.5cm 0.0cm 0.4cm}, clip, width=0.48\textwidth]{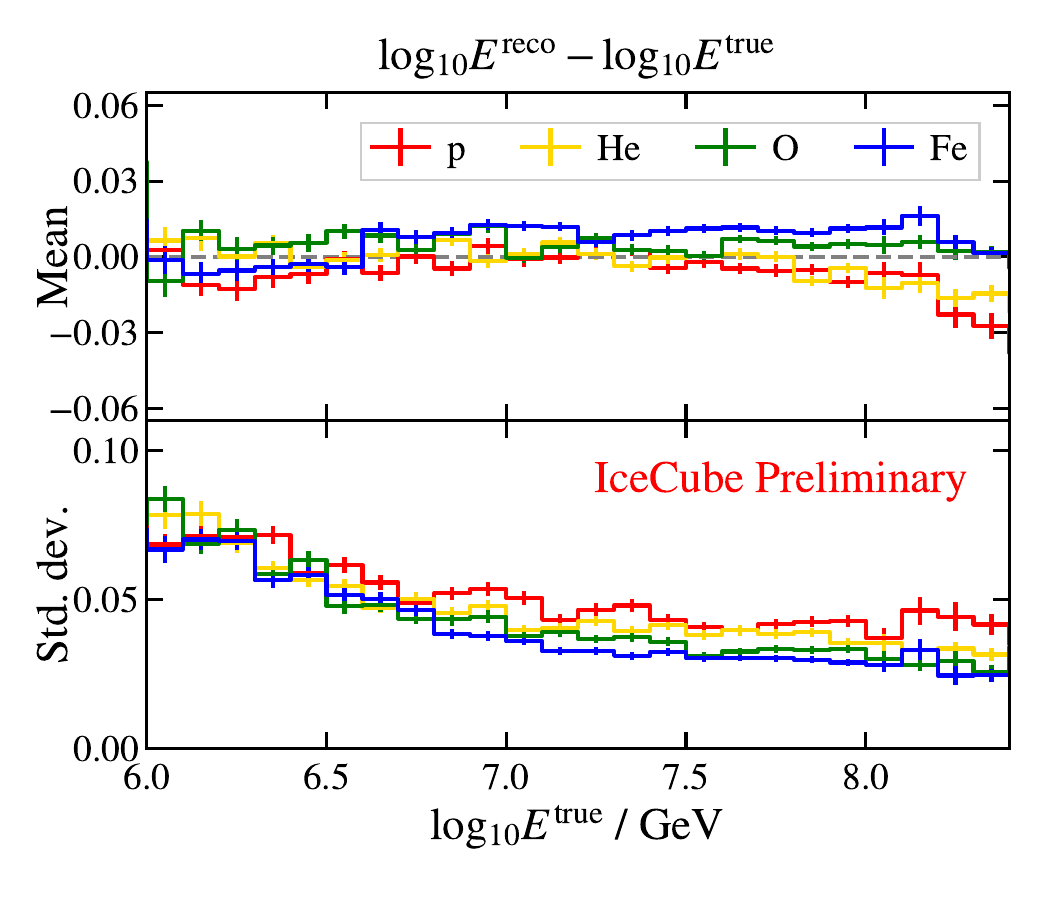}\hfill\includegraphics[trim={0.0cm 0.5cm 0.0cm 0.4cm}, clip, width=0.48\textwidth]{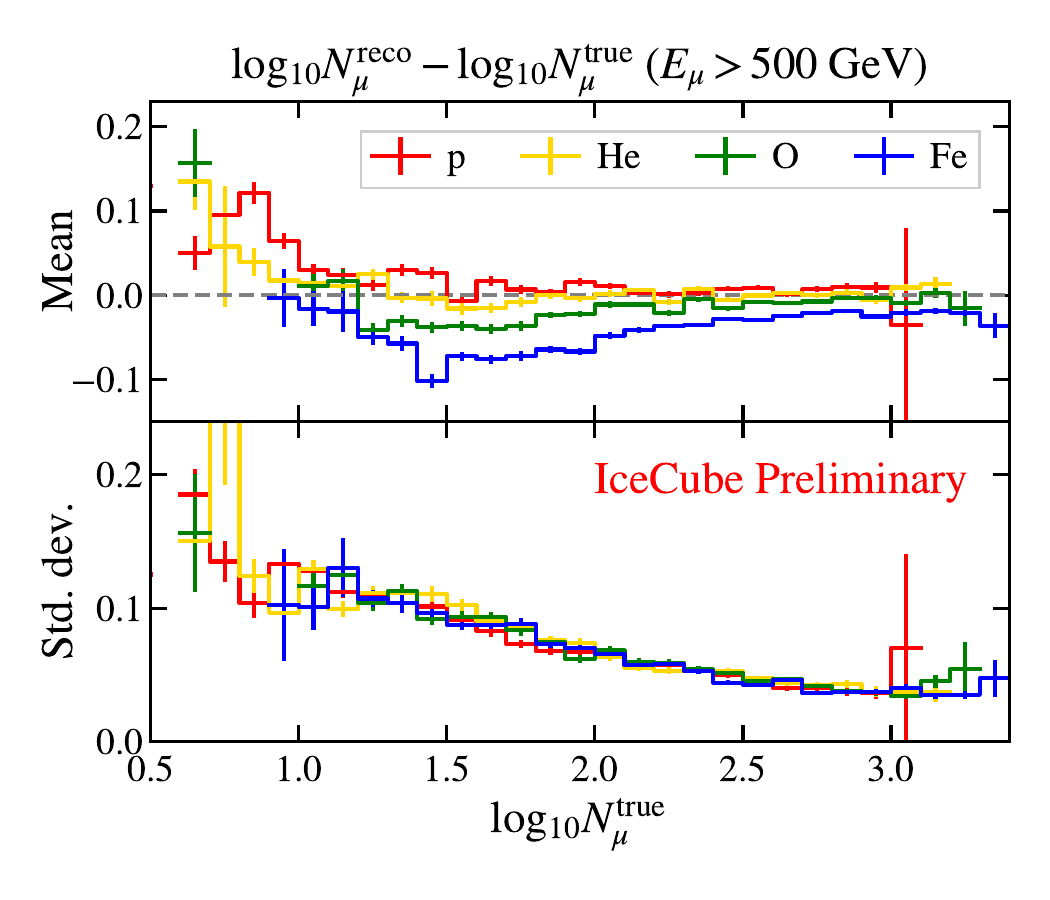}
    \caption{Bias and resolution of the neural network trained in this analysis, defined as the the mean and standard deviation of the difference between the logarithms of the reconstructed and true values. Left: Primary energy \Enul{}. Right: Muon multiplicity \Nmu{} (for $E_\mu > \SI{500}{\giga\eV}$).}
    \label{fig:performance}
\end{figure}

\reffig{fig:performance} shows the performance of the neural network reconstructions of \Enul{} and \Nmu{} in terms of bias and resolution as a function of the true values for the different primary nuclei, defined as the mean and standard deviation of a Gaussian fit to the difference between the $\log_{10}$ of true and reconstructed values.

\subsection{Mean muon number: MC correction}\label{sec:correction}

\begin{figure}
    \centering
    \includegraphics[trim={0.3cm 0.5cm 0.3cm 0.4cm}, clip, width=0.5\textwidth]{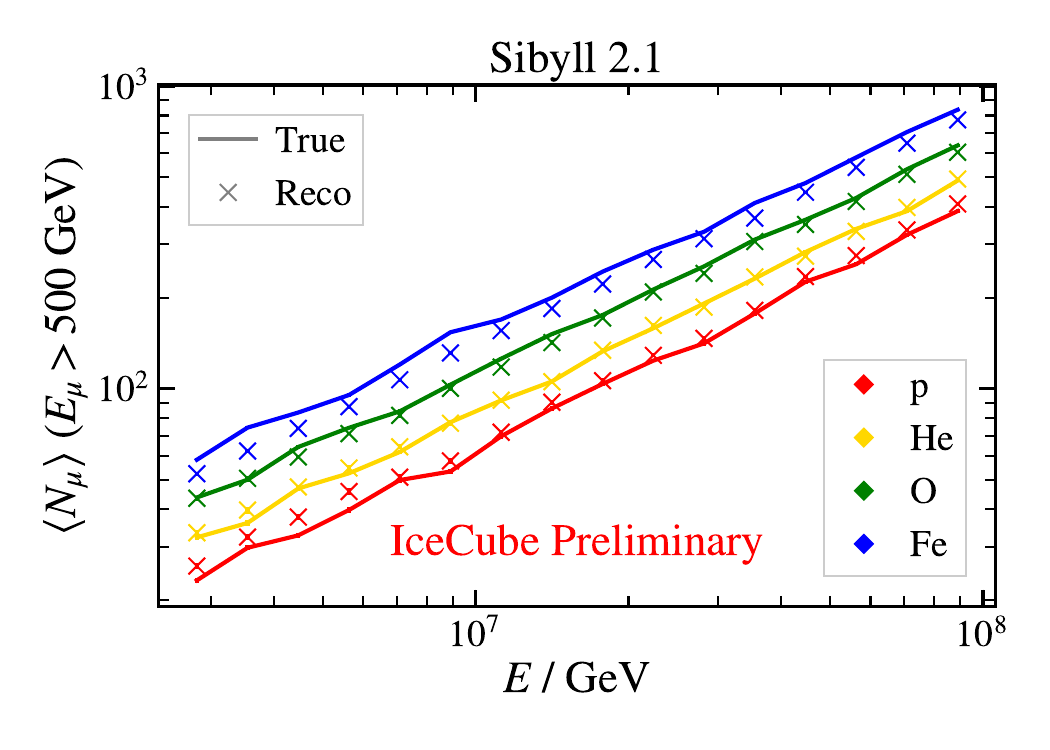}\includegraphics[trim={0.3cm 0.5cm 0.3cm 0.4cm}, clip, width=0.5\textwidth]{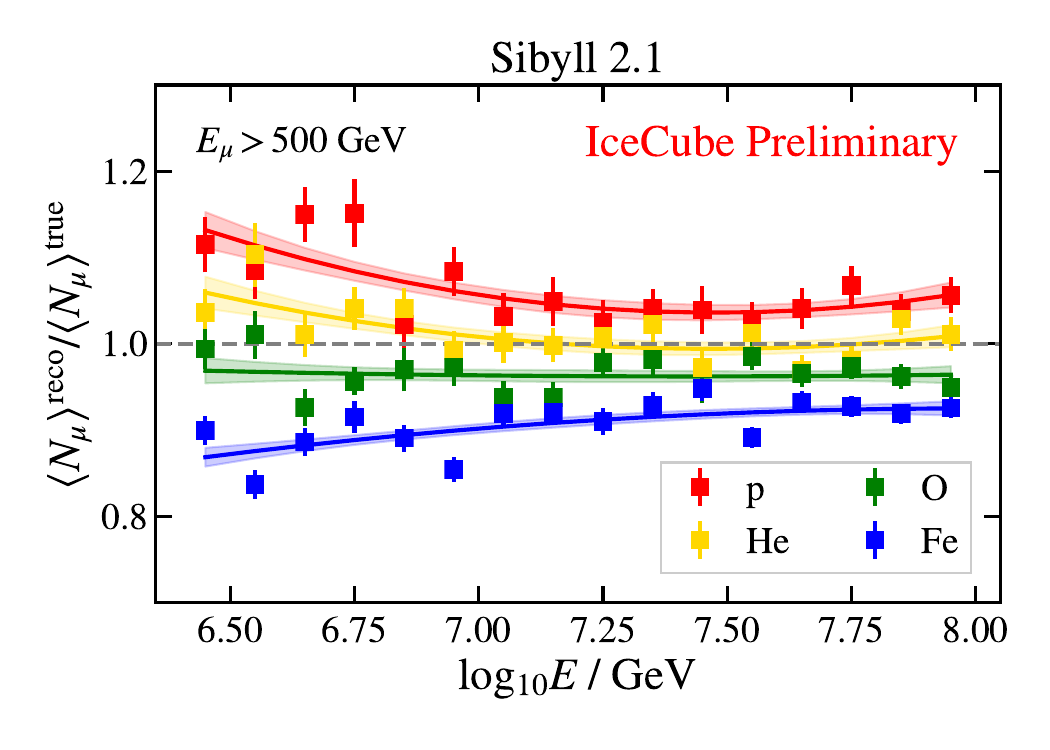}
    \caption{Left: Comparison of the average reconstructed high-energy muon multiplicity \Nmuav{} in bins of reconstructed cosmic-ray energy \Enul{} and the true values in simulation for four different primary types. Right: Ratio of the reconstructed and true \Nmuav{} from the left panel. Biases of about $\pm 15\%$ depending on the primary are observed. The ratios are fitted with quadratic functions, which are used further in the analysis as multiplicative correction factors.}
    \label{fig:correction}
\end{figure}

The average muon multiplicity as function of primary energy is estimated by binning events in the neural-network reconstructed \Enul{} and calculating the mean reconstructed \Nmu{}. The accuracy of the method can be obtained by performing it on simulation and comparing to the true mean \Nmu{} in bins of true \Enul{}. Such a comparison is shown in \reffig{fig:correction} (left). Although there is good agreement in the general behavior, systematic biases are present. The magnitude of the bias depends on both energy and mass. The ratio of the true and reconstructed values is fitted with quadratic functions, as shown in \reffig{fig:correction} (right). These fits are used as correction factors which will be applied to the results obtained from data.

\begin{wrapfigure}{R}{0.5\textwidth}
    \centering
    \includegraphics[trim={0.5cm 0.5cm 0.5cm 0.4cm}, clip, width=0.5\textwidth]{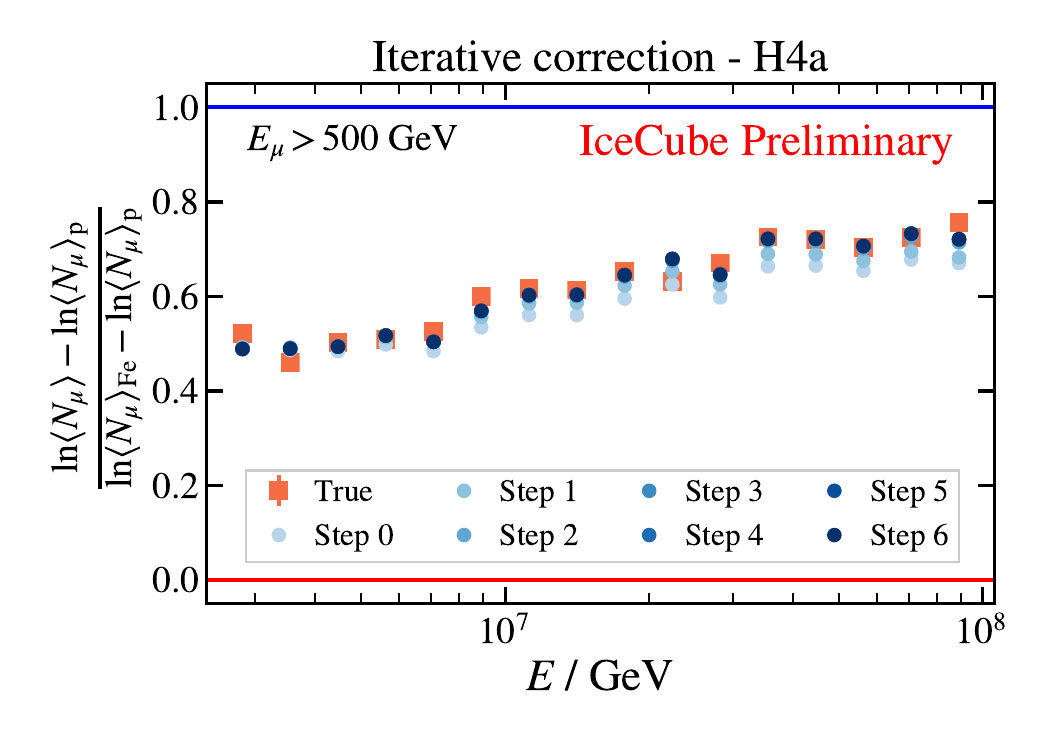}
    \caption{Example of the iterative correction procedure applied to simulation weighted according to the H4a composition model. ``Step 0'' shows \Nmuav{} obtained after applying the neural network reconstructions and obtaining the average value in each bin. After several steps of applying the correction, updating the correction factor in each step as described in \refsec{sec:correction}, \Nmuav{} can be seen to converge to values consistent with the true values in the simulation.\vspace{-0.4cm}}
    \label{fig:iterative}
\end{wrapfigure}

One complication is the mass dependence of the correction factors. To handle this without having to assume a specific mass composition model, an iterative method is used. It takes advantage of the approximately linear dependence of the correction factors on $\ln A$, where $A$ is the mass number of the nucleus, visible as the approximate equidistance of the four fits at each energy in the right panel of \reffig{fig:correction}. To accomplish this, the average \Nmuav{} obtained in an energy bin is compared to predictions for proton and iron simulation using
\begin{equation}\label{eq:z}
    z = \frac{\ln \langle N_\mu \rangle - \ln \langle N_\mu \rangle_\mathrm{p}}{\ln \langle N_\mu \rangle_\mathrm{Fe} - \ln \langle N_\mu \rangle_\mathrm{p}},
\end{equation}
also referred to as ``$z$-scale''~\cite{EAS-MSU:2019kmv}. The Heitler-Matthews model and the superposition principle predict that a muon measurement corresponds to an estimate of the mass composition, $z \approx \ln A / \ln 56$~\cite{Matthews:2005sd}. The fitted correction factors of \reffig{fig:correction} for proton and iron, $\mathcal{C}_\mathrm{p}$ and $\mathcal{C}_\mathrm{Fe}$, are linearly interpolated to find the correction factor corresponding to the reconstructed $z$, 
\begin{equation}
\mathcal{C}(\ln A) = \mathcal{C}_\mathrm{p} + \frac{\mathcal{C}_\mathrm{Fe}-\mathcal{C}_\mathrm{p}}{\ln 56}\ln A.
\end{equation}
The correction factor obtained in this way is then applied to the initial \Nmuav{}. With the updated estimate of \Nmuav{}, a new correction factor can be constructed. This process is repeated until the values of \Nmuav{} converge. An example of this iterative correction procedure applied to simulations is shown in \reffig{fig:iterative}. It has been tested that this approach successfully obtains \Nmuav{} values compatible with true values, regardless of the underlying mass composition. Note that because of how the correction factors are derived, they also implicitly correct for biases which result from events migrating to different energy bins.
In addition to the correction factors derived from \sibyllpre{} simulations, correction factors were also derived by applying the neural network to \qgsjet{} and \epos{} simulations. In this way, results can be derived from experimental data under different model assumptions.

\subsection{Systematic uncertainties}

The systematic uncertainties in this analysis arise from several sources. The detector uncertainties are those described in \refref{IceCube:2019hmk}, however, with a more conservative value of 10$\%$ for the uncertainty on the DOM efficiency. Other uncertainties relevant for the in-ice detector are related to the scattering and absorption of Cherenkov photons in the Antarctic ice. These are the dominant uncertainties on the final result. For IceTop, the uncertainties are related to the snow accumulation on the detector and the calibration of the charge unit (vertical equivalent muon or VEM).

The statistical uncertainties derived on the correction factors (see \reffig{fig:correction}) are also included as a systematic uncertainty on the final result.

\section{Results}

\begin{figure}
    \centering
    \includegraphics[trim={0.3cm 0.5cm 0.3cm 0.4cm}, clip, width=0.5\textwidth]{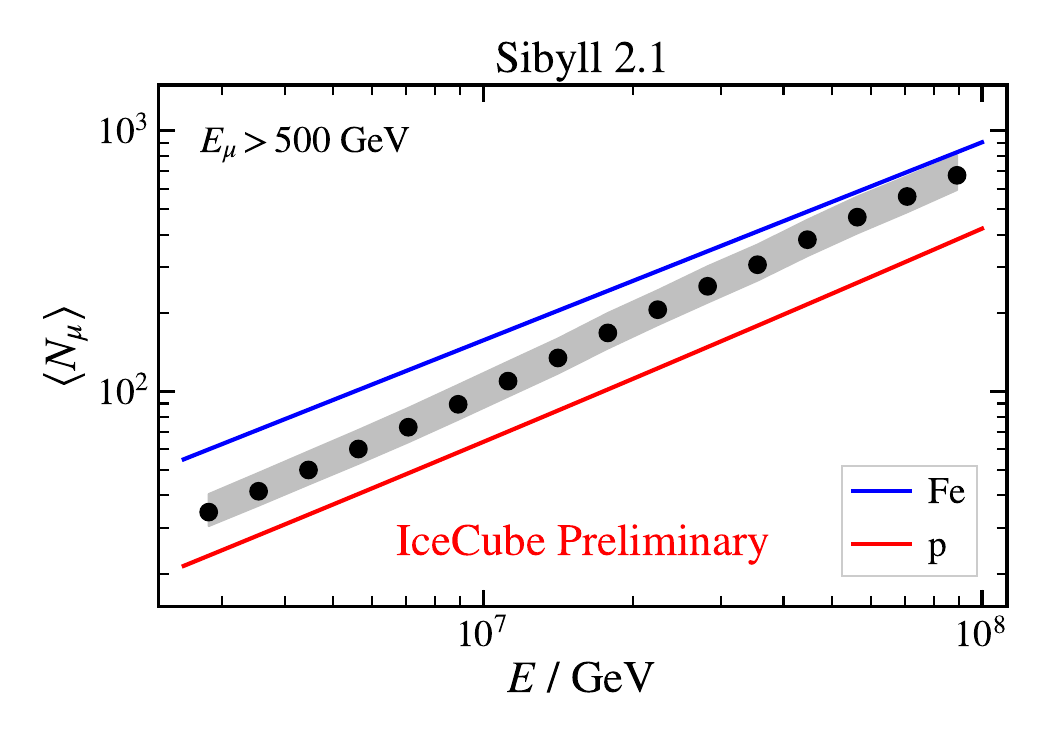}\includegraphics[trim={0.3cm 0.5cm 0.3cm 0.4cm}, clip, width=0.5\textwidth]{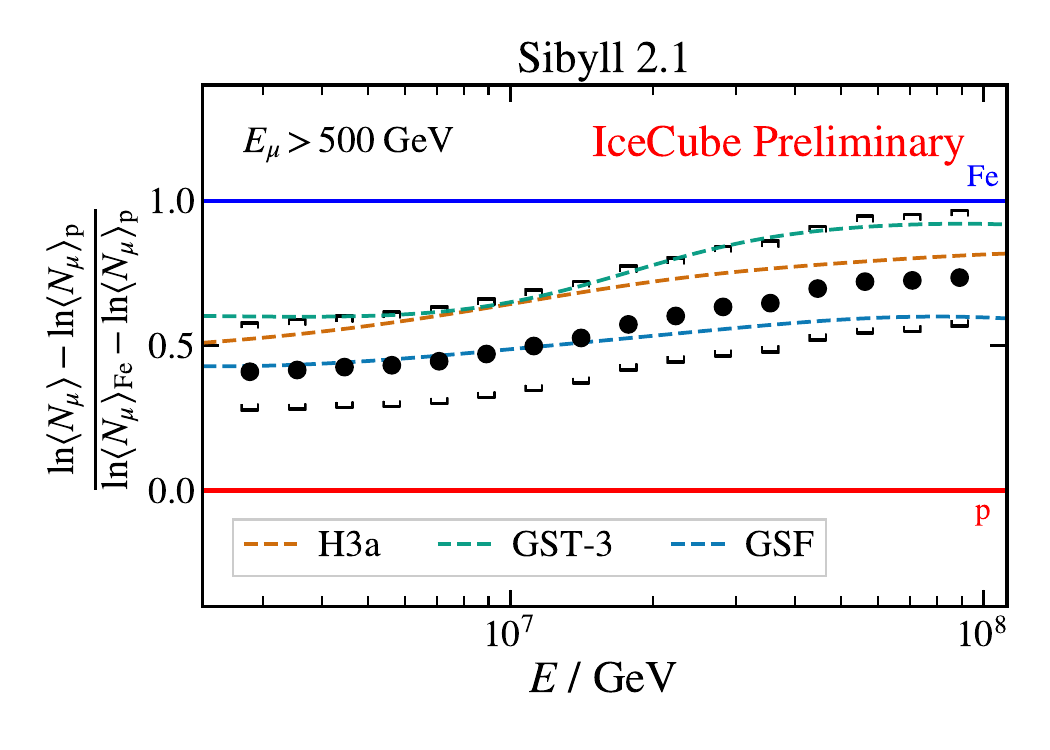}

    \includegraphics[trim={0.3cm 0.5cm 0.3cm 0.4cm}, clip, width=0.5\textwidth]{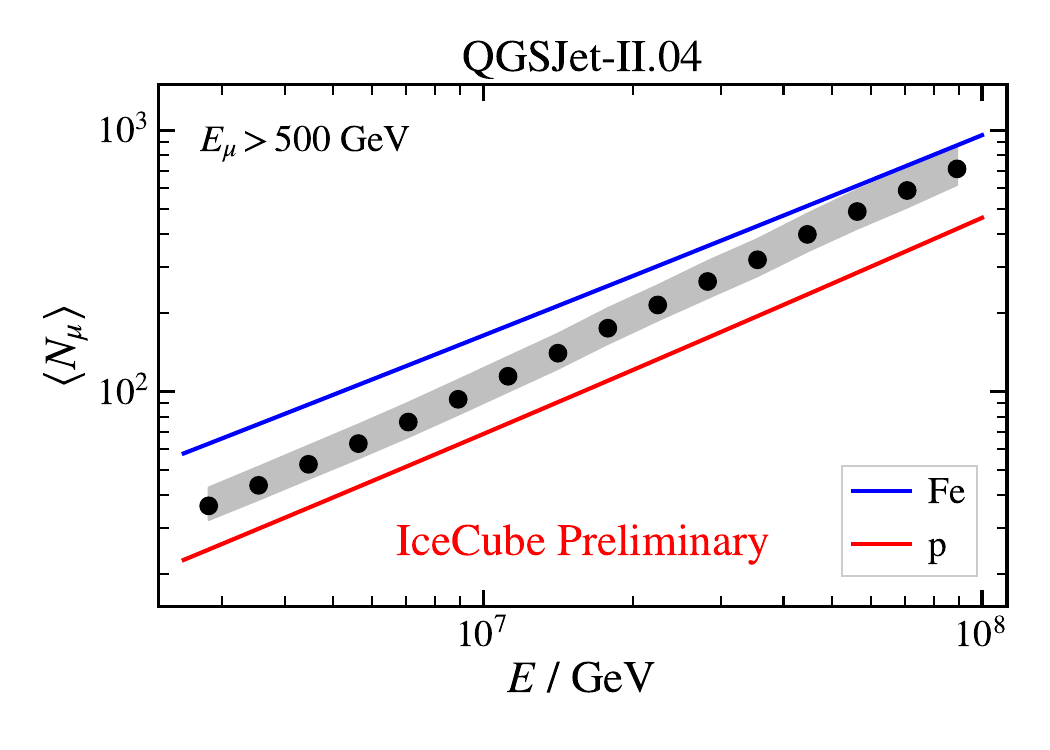}\includegraphics[trim={0.3cm 0.5cm 0.3cm 0.4cm}, clip, width=0.5\textwidth]{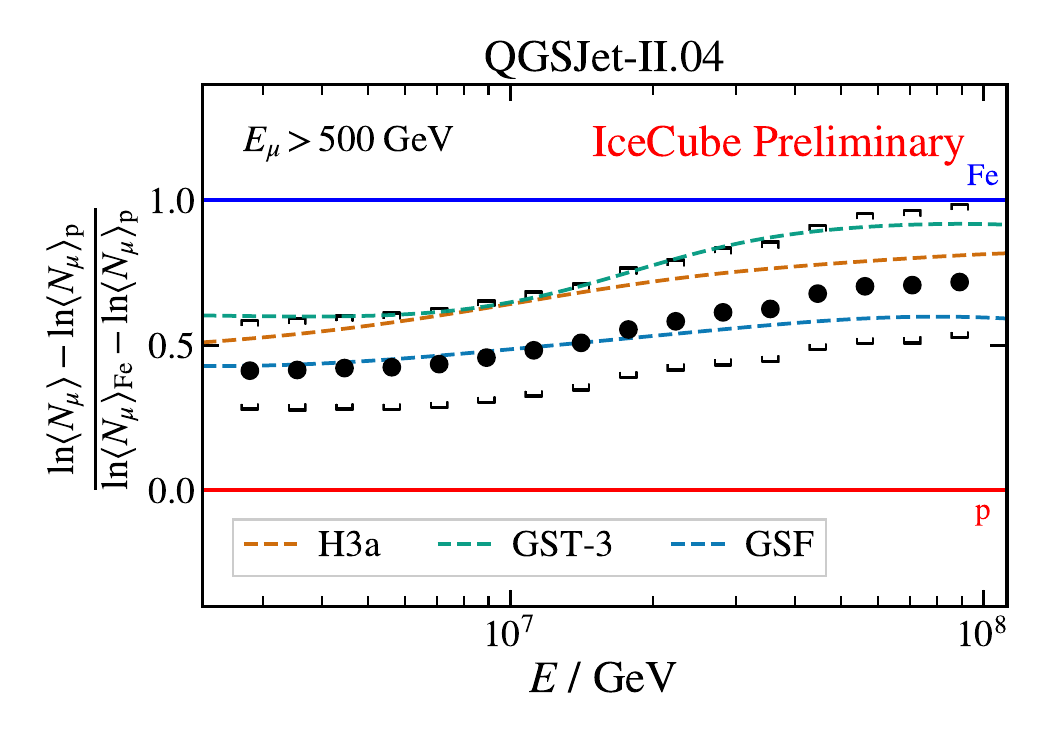}

    \includegraphics[trim={0.3cm 0.5cm 0.3cm 0.4cm}, clip, width=0.5\textwidth]{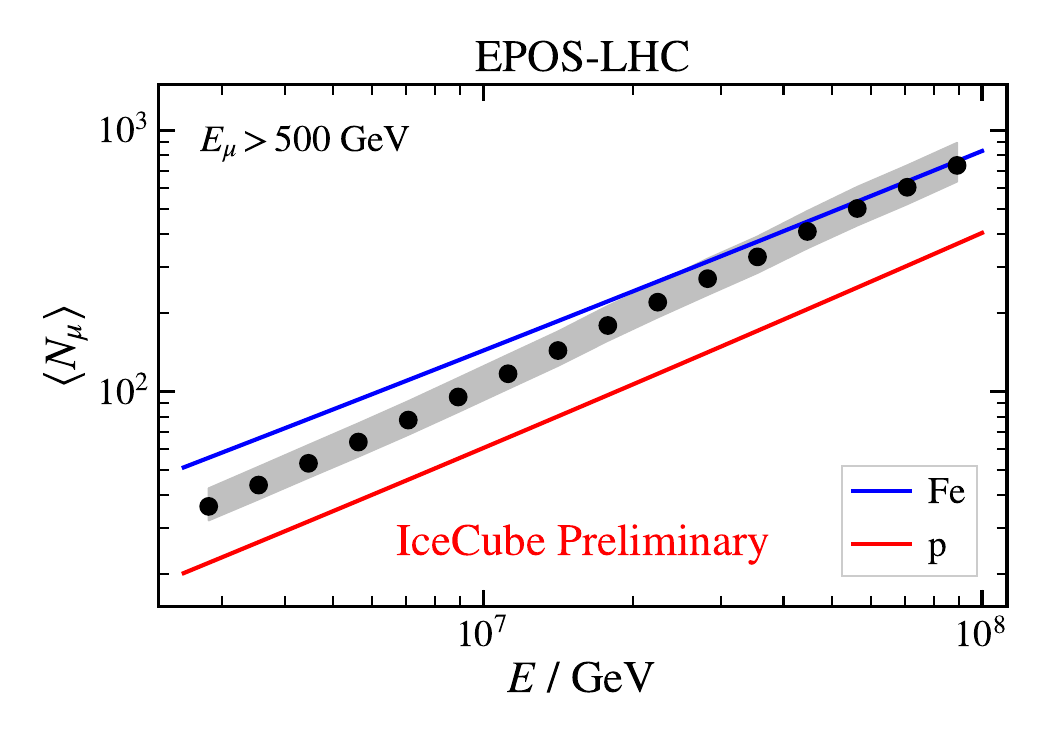}\includegraphics[trim={0.3cm 0.5cm 0.3cm 0.4cm}, clip, width=0.5\textwidth]{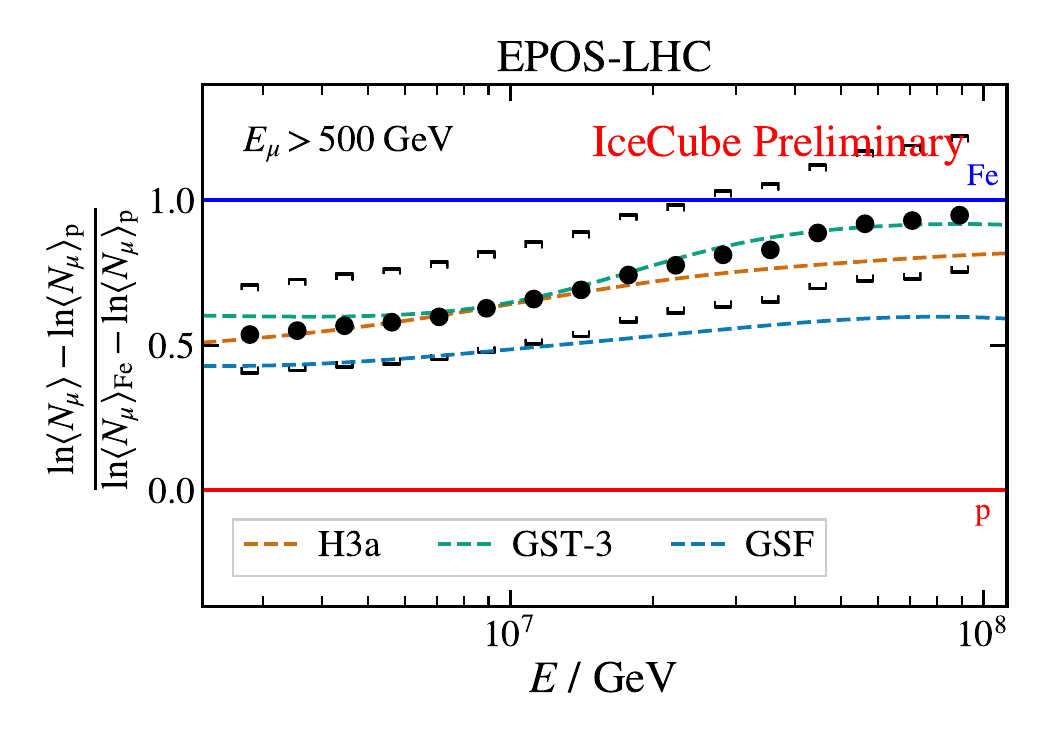}
    \caption{Average number of muons with energy above \SI{500}{\giga\eV} as a function of primary energy for near-vertical EAS, obtained from experimental data using the hadronic interaction models \sibyllpre{}, \qgsjet{}, and \epos{}. Plots also show predictions from proton and iron simulations based on the same models. The figures on the right show the corresponding z-values (\refeq{eq:z}), with expectations from cosmic-ray flux models for comparison. Bands and brackets indicate the total systematic uncertainty, error bars are plotted for the statistical uncertainties but are smaller than the marker size.}
    \label{fig:results}
\end{figure}

The average multiplicity of muons above \SI{500}{\giga\eV}, derived from one year of experimental data under the assumption of the hadronic interaction models \sibyllpre{}, \qgsjet{}, and \epos{}, is presented in \reffig{fig:results} (left). The results for each model are shown together with predictions from simulations based on the corresponding model. In order to visualize the composition implied by the measurements in more detail, they are plotted by scaling them according to the predictions, follwing \refeq{eq:z}. This is shown in \reffig{fig:results} (right), compared to expectations from the cosmic-ray flux models GSF~\cite{Dembinski:2017zsh}, GST~\cite{Gaisser:2013bla}, and H3a~\cite{Gaisser:2011klf}. Although the result obtained using \epos{} indicates a composition which is consistently heavier than those obtained with \sibyllpre{} and \qgsjet{}, all results are in agreement with expectations from simulations within uncertainties.

It is of interest to compare the high-energy muon result presented here to the GeV muon density measurement performed with IceTop alone in the same primary energy range, as presented in \refref{IceCubeCollaboration:2022tla}. The composition interpretation of the GeV and TeV muon results should be consistent if the simulations give an accurate description of EAS. This is the case for the results obtained with \sibyllpre{}. For the post-LHC models \qgsjet{} and \epos{}, however, a tension is observed as a result of their increased number of GeV muons compared to \sibyllpre{}, indicating a very light mass composition. In addition, indications of a discrepancy between the muon measurements for \sibyllpre{} and a shower observable related to the IceTop LDF have been observed in \refref{IceCube:2021ixw}.

\section{Conclusion}\label{sec3}

We have presented a first measurement of the multiplicity of TeV muons in near-vertical EAS with energies ranging from \SI{2.5}{\peta\eV} to \SI{100}{\peta\eV} with the IceTop and IceCube detectors. The results agree within uncertainties with predictions from simulations for all hadronic interaction models considered, i.e. \sibyllpre{}, \qgsjet{}, and \epos{}. Discrepancies are observed with the measurement of the GeV muon density with IceTop for \qgsjet{} and \epos{}, suggesting these models do not adequately describe the experimental data in this energy range.

Future improvements in this analysis will decrease the systematic uncertainties related to the ice model~\cite{Rongen:2021rgc} and push the analysis to higher primary energies. Together with improvements in low-energy muon measurements, such as the event-by-event approach explored in \refref{weyrauch:2023icrc}, the spectral information obtained from combined muon measurements with IceTop and IceCube will be able to provide even more stringent constraints on muon production in EAS~\cite{Riehn:2019jet}. An improved description of high-energy hadronic interactions is not only crucial for the interpretation of cosmic-ray measurements, but also benefits other domains of particle astrophysics, such as measurements of astrophysical neutrinos and gamma rays~\cite{Coleman:2022abf}.


\bibliographystyle{ICRC}
\bibliography{references}

\clearpage

\section*{Full Author List: IceCube Collaboration}

\scriptsize
\noindent
R. Abbasi$^{17}$,
M. Ackermann$^{63}$,
J. Adams$^{18}$,
S. K. Agarwalla$^{40,\: 64}$,
J. A. Aguilar$^{12}$,
M. Ahlers$^{22}$,
J.M. Alameddine$^{23}$,
N. M. Amin$^{44}$,
K. Andeen$^{42}$,
G. Anton$^{26}$
% ...
\\
\\
% ...
$^{12}$ Universit{\'e} Libre de Bruxelles, Science Faculty CP230, B-1050 Brussels, Belgium \\
% ...
$^{17}$ Department of Physics, Loyola University Chicago, Chicago, IL 60660, USA \\
$^{18}$ Dept. of Physics and Astronomy, University of Canterbury, Private Bag 4800, Christchurch, New Zealand \\
% ...
$^{22}$ Niels Bohr Institute, University of Copenhagen, DK-2100 Copenhagen, Denmark \\
$^{23}$ Dept. of Physics, TU Dortmund University, D-44221 Dortmund, Germany \\
% ...
$^{26}$ Erlangen Centre for Astroparticle Physics, Friedrich-Alexander-Universit{\"a}t Erlangen-N{\"u}rnberg, D-91058 Erlangen, Germany \\
% ...
$^{40}$ Dept. of Physics and Wisconsin IceCube Particle Astrophysics Center, University of Wisconsin{\textendash}Madison, Madison, WI 53706, USA \\
% ...
$^{42}$ Department of Physics, Marquette University, Milwaukee, WI, 53201, USA \\
% ...
$^{44}$ Bartol Research Institute and Dept. of Physics and Astronomy, University of Delaware, Newark, DE 19716, USA \\
% ...
$^{63}$ Deutsches Elektronen-Synchrotron DESY, Platanenallee 6, 15738 Zeuthen, Germany  \\
$^{64}$ Institute of Physics, Sachivalaya Marg, Sainik School Post, Bhubaneswar 751005, India \\
% ...

\subsection*{Acknowledgements}

\noindent
The authors gratefully acknowledge the support from the following agencies and institutions:
USA {\textendash} U.S. National Science Foundation-Office of Polar Programs,
U.S. National Science Foundation-Physics Division,
U.S. National Science Foundation-EPSCoR,
Wisconsin Alumni Research Foundation,
Center for High Throughput Computing (CHTC) at the University of Wisconsin{\textendash}Madison,
Open Science Grid (OSG),
Advanced Cyberinfrastructure Coordination Ecosystem: Services {\&} Support (ACCESS),
Frontera computing project at the Texas Advanced Computing Center,
U.S. Department of Energy-National Energy Research Scientific Computing Center,
Particle astrophysics research computing center at the University of Maryland,
Institute for Cyber-Enabled Research at Michigan State University,
and Astroparticle physics computational facility at Marquette University;
Belgium {\textendash} Funds for Scientific Research (FRS-FNRS and FWO),
FWO Odysseus and Big Science programmes,
and Belgian Federal Science Policy Office (Belspo);
Germany {\textendash} Bundesministerium f{\"u}r Bildung und Forschung (BMBF),
Deutsche Forschungsgemeinschaft (DFG),
Helmholtz Alliance for Astroparticle Physics (HAP),
Initiative and Networking Fund of the Helmholtz Association,
Deutsches Elektronen Synchrotron (DESY),
and High Performance Computing cluster of the RWTH Aachen;
Sweden {\textendash} Swedish Research Council,
Swedish Polar Research Secretariat,
Swedish National Infrastructure for Computing (SNIC),
and Knut and Alice Wallenberg Foundation;
European Union {\textendash} EGI Advanced Computing for research;
Australia {\textendash} Australian Research Council;
Canada {\textendash} Natural Sciences and Engineering Research Council of Canada,
Calcul Qu{\'e}bec, Compute Ontario, Canada Foundation for Innovation, WestGrid, and Compute Canada;
Denmark {\textendash} Villum Fonden, Carlsberg Foundation, and European Commission;
New Zealand {\textendash} Marsden Fund;
Japan {\textendash} Japan Society for Promotion of Science (JSPS)
and Institute for Global Prominent Research (IGPR) of Chiba University;
Korea {\textendash} National Research Foundation of Korea (NRF);
Switzerland {\textendash} Swiss National Science Foundation (SNSF);
United Kingdom {\textendash} Department of Physics, University of Oxford.

\end{document}